\documentclass[doublecol,figures]{epl2}
\usepackage[colorlinks,citecolor=blue,urlcolor=blue,linkcolor=blue]{hyperref}
\usepackage{bm}% bold math
\usepackage{graphicx}
\usepackage{latexsym,color}
\usepackage{amssymb}
\newcommand{\ti}[1]{\mbox{\tiny{#1}}}

\newcommand{\be}{\begin{equation}}
\newcommand{\ee}{\end{equation}}
\newcommand{\bea}{\begin{eqnarray}}
\newcommand{\eea}{\end{eqnarray}}

\newcommand{\il}{~}

\newcommand{\Sie}{\mathcal{S}}
\newcommand{\Mie}{\mathcal{M}}

\title{Squeezing of  toroidal accretion disks}
\author{D. Pugliese\inst{1}\thanks{E-mail: \email{daniela.pugliese@alice.it}} \and G. Montani\inst{2,3}\thanks{E-mail: \email{giovanni.montani@frascati.enea.it}}}
\shortauthor{D. Pugliese \etal}

\institute{
  \inst{1}School of Mathematical Sciences, Queen Mary University of London - Mile
End Road, London E1 4NS, UK, EU\\
  \inst{2} ENEA - C.R, UTFUS-MAG -
Via Enrico Fermi 45, 00044
Frascati, Roma, Italy, EU\\
\inst{3} Dipartimento di Fisica, Universit\`a di Roma ``Sapienza"- Piazzale Aldo Moro 5, I-00185 Roma, Italy, EU
}

\date{\today}% The correct dates will be entered by the editor

\abstract{
Accretion disks around very compact objects such as very massive Black hole can grow according to thick toroidal models.
We face the problem of defining how  does change the thickness of a toroidal accretion disk  spinning around a Schwarzschild Black hole under the influence of  a toroidal magnetic field and by varying the fluid  angular momentum. We consider both an hydrodynamic and  a magnetohydrodynamic disk  based on the Polish doughnut thick model. We show that the torus thickness remains basically unaffected but tends to increase or decrease slightly  depending on the balance of the magnetic, gravitational and centrifugal effects   which the disk is subjected to.
}
\begin{document}
\maketitle
\section{Introduction}
Configurations of extended matter rotating around gravitational sources  are undoubtedly an environment  with  a rich phenomenology covering very different aspects of Astrophysics in different energetic sectors, from for example the proto-planetary accretion disks  to, most likely,  violent dynamical effects  such as Gamma Ray Bursts (GRB), see e.g.\cite{FeB04,So07}.
 In  addition,  since many of these disks are made of plasma and they share their toroidal topology with other plasma devices in the terrestrial physic laboratory , for example Tokamak machines, the questions about  the plasma stability and  the confinement under magnetic field, involve in general a much wider interest of the astrophysic phenomenology.
There are  different ways to classify accretion disk models, for example according to optical depth, accretion rates and geometric structure: in general accretion disks have toroidal shape, but
we can roughly distinguish between thick  and thin disk models,
in this work we consider a thick toroidal disk. The dynamics of any accretion disk is determined by several factors, say  centrifugal, dissipative, magnetic effects.
In a sense we could say that the thick accretion disks are strongly characterized by the  gravitational ones in the following respects: thick models are supposed to require attractors that are very compact sources of strong gravitational fields, for example Black hole and  they characterize the physics of most energetic  astrophysical objects as Active Galactic Nuclei  (AGN) or GRB. Since they characterize  regions very close to the source, it is   in many cases necessary a full general relativistic treatment of the model. Furthermore  in many thick models gravitational force constitutes the basic ingredient of the accretion mechanism   independently of any dissipative effects that are   strategically   important   for the accretion process in the thin models \cite{Balbus2011,Shakura1973}.
The  Polish doughnut (PD) is an
{opaque}  (large optical depth) and
Super-Eddington (high  matter accretion rates) disk, characterized by
an
ad hoc distributions of constant angular momentum. This model and its derivations are widely studied in the literature with both numerical and analytical methods, we refer  for  an extensive bibliography to \cite{Abramowicz:2011xu,Pugliese:2012ub}.
The torus shape is defined by the constant Boyer potential $W$:
closed equipotential surfaces define stationary equilibrium configurations, the fluid
can fill any closed surface.
The open equipotential surfaces define
dynamical situations, for example the formation of matter jets \cite{Boy:1965:PCPS:}.
The critical, self-crossing and closed equipotential surfaces (cusp) locates the accretion onto the Black hole due to Paczy\'nski
mechanism:
a little overcoming of the critical equipotential surface, violation of the hydrostatic equilibrium when
  the disk surface exceeds
the critical equipotential surface $W_{cusp}$.
The relativistic
Roche lobe overflow at the cusp of  the equipotential surfaces is also  the
 stabilizing mechanism against the thermal and viscous instabilities locally,
and against the so called Papaloizou and Pringle instability globally \cite{Blaes1987}.
In  \cite{Komissarov:2006nz} and \cite{Montero:2007tc},  the fully relativistic theory of
stationary axisymmetric  PD torus in Kerr metric has been generalized   by including
a strong toroidal magnetic field,  leading to the  analytic solutions for barotropic tori with constant
angular momentum.
In a recent work  the  PD model has been revisited providing  a  comprehensive
analytical description of the PD structure, considering  a perfect  fluid circularly orbiting  around a Schwarzschild background geometry, by the effective potential approach for the exact gravitational and centrifugal
effects \cite{Pugliese:2012ub}.
The work has taken massively advantage of  the structural analogy of the potential for the fluid with the corresponding function for the free particle dynamics  which is not subjected to kinetic pressure   in the  same orbital configuration around the source.
This analysis leads to a  detailed, analytical description of the accretion
disk, its toroidal surface, the thickness, the  distance from the source  with the effective potential and the fluid angular
momentum. A series of results concerning the PD shape was found, in particular it was   proved  that the torus thickness  $(h)$ increases with the
 ``energy'' parameter $({\rm{K}})$, defined as the
value of the effective potential for that momentum, and decreases with the angular momentum: the torus becomes thinner for
    high angular momenta, and  thicker  for high energies and
the { maximum diameter} $(\lambda)$ increases with ${\rm{K}}$, but
    decreases with the fluid angular momentum, moreover the location of maximum thickness of the torus moves  towards the external
   regions  with increasing angular momentum and energy, until it reaches a maximum    an then decreases.

In this paper we generalize this approach to the   PD disk to  the  magnetohydrodynamic models developed in \cite{Komissarov:2006nz,Montero:2007tc}. We address  the specific question of torus squeezing on the equatorial plane  exploring the disk thickness changing  the physical characteristics of the torus  as the angular momentum and the effective potential  which characterize the  toroidal surface, and the magnetic field for the magnetohydrodynamic case, then we will find the extreme squeezing.
\section{Fluid configuration}\label{Sec:fluidConf}
We  consider  the motion  in the  Schwarzschild background
geometry written as
\begin{equation}\label{11metrica}
ds^2=-e^{\nu(r)}dt^2+e^{-\nu(r)}(r)dr^2
+r^2\left(d\theta^2+\sin^2\theta d\phi^2\right),
\end{equation}
in the spherical polar coordinate $(t,r,\theta,\phi)$ where $e^{\nu(r)}\equiv\left(1-2M/r\right)$.
We  define $
\Lambda\equiv U^r$, $\Sigma\equiv U^t$, $\Phi\equiv U^{\phi}$, $\Theta\equiv U^{\theta}$,
for the fluid  four-velocity vector field  $U$,
and we introduce the  set  of variables $\{E, V_{sc}, L, T\}$  by the following
relations\footnote{We adopt the
geometrical  units $c=G=M=1$ extended to the electromagnetic quantities
by setting $\epsilon_0 =1$, where $\epsilon_0$ is the vacuum permittivity and  the $(-,+,+,+)$ signature.}:
\bea\label{La2fluido}
\Lambda =\sqrt{E^2-V_{sc}^2}\, ,\quad
\Sigma =\frac{E}{e^{\nu}}\, ,\quad
\Phi = \frac{L}{r^2\sin^2\theta}\, ,\quad
\Theta=\frac{T}{r^2}\, ,
\eea
where
\be\label{60}
V_{sc}\equiv\sqrt{e^{\nu(r)}\left(1+\frac{L^2}{r^2\sin^2\theta}+\frac{T^2}{r^2}\right)}
\ee
is the \emph{effective potential} \cite{Pugliese:2012ub}.
We consider the case of a fluid circular configuration, defined by the constraints
$\Lambda=0$ (i.e. ${V_{sc}}=E$), restricted to a fixed plane $\sin\theta=\sigma\neq0$. No
motion is assumed in the $\theta$ angular direction, which means $\Theta=0$. For the
symmetries of the problem, we always assume $\partial_t \mathbf{Q}=0$ and
$\partial_{\phi} \mathbf{Q}=0$, being $\mathbf{Q}$ a generic tensor of the spacetime.
The fluid is characterized by the constant angular momentum:
$$
\ell\equiv\frac{g_{\phi\phi}}{g_{tt}}\frac{\Phi}{\Sigma}=\frac{L}{V_{sc}}.$$

We consider  an infinitely conductive plasma: $F_{ab}U^a=0$, where $F_{ab}$ is the Faraday tensor.
Using the equation $U^a B_a=0$, where $B^a$ is the magnetic field,  we obtain the relation  $B^t=\Phi B^{\phi}/\Sigma$.
We note that, if we set $B^r=0$ from the Maxwell equations, we infer
\(
B^{\theta}\cot\theta=0
\)
(with $\ell=$constant), that is satisfied for $B^{\theta}=0$ or $\theta=\pi/2$.
The former implies that the assumptions $\partial_{\phi}  B^{\phi}=0$ and $B^r=0$ together lead to $B^{\theta}=0$,
 in this work we consider $\partial_{\phi}B^a=0$ and $B^r=B^{\theta}=0$.
As noted in \cite{Komissarov:2006nz} the presence of a magnetic field with a predominant toroidal component is  reduced to the disk differential rotation, viewed as a generating  mechanism of the magnetic field,  for further discussion we refer  to \cite{Komissarov:2006nz,Montero:2007tc,Horak:2009iz} and
 \cite{Parker:1955zz,Parker:1970xv,Y.I.I2003,R-ReS1999}.

The Euler equation for this system has been exactly integrated for the background spacetime of Schwarzschild and Kerr Black
holes in  \cite{Komissarov:2006nz,Montero:2007tc,Horak:2009iz} assuming   the magnetic field is
\be\label{RSC}
B^{\phi }=\sqrt{\frac{2 p_B}{g_{\phi \phi }+2 \ell  g_{t\phi}+\ell ^2g_{tt}}},
\ee
where
\(
p_B=\mathcal{M} \left(g_{t \phi }g_{t \phi }-g_{{tt}}g_{\phi \phi }\right){}^{q-1}\omega^q
\) is the magnetic pressure,
$\omega$ is the fluid enthalpy, $q$  and $\Mie$ are constant, we assume moreover a barotropic equation of state.
We can write the Euler equation   as
\bea\label{due}
\frac{\partial_rp}{\rho+p}&=&G_r^{(f)}+G_r^{(em)},\quad
\frac{\partial_{\theta}p}{\rho+p}=G_{\theta}^{(f)}+G_{\theta}^{(em)},
\eea
 where
using eq.\il(\ref{RSC}), it follows
\bea
G_{\theta}^{(em)}&=&-\frac{\partial}{\partial\theta}W^{(em)}_{\theta},\quad G_{r}^{(em)}=-\frac{\partial}{\partial r}W^{(em)}_{r},
\\
G_{\theta}^{(f)}&=&-\frac{\partial}{\partial\theta}W^{(f)}_{\theta},\quad G_{r}^{(f)}=-\frac{\partial}{\partial r}W^{(f)}_{r},
\eea
with
\bea
W_{r}^{(f)}&\equiv&\ln V_{sc},\quad W_{r}^{(em)}\equiv \mathcal{G}_r(r,\theta)+g_{r}(\theta),
\\
W_{\theta}^{(f)}&\equiv&\ln V_{sc},\quad W_{\theta}^{(em)}\equiv \mathcal{G}_{\theta}(r,\theta)+g_{\theta}(r),
\eea
where $g_{\theta}(r)$ and $g_{r}(\theta)$ are functions to be fixed by the integration.
For $q\neq1$ it is
\(
\mathcal{G}_r(r,\theta)=\mathcal{G}_{\theta}(r,\theta)=\mathcal{G}(r,\theta),
\)
where
\bea
\mathcal{G}(r,\theta)&\equiv& \frac{\Mie q \left[(r-2) r \sigma^2\right]^{(q-1)} \omega^{(q-1)}}{q-1}.
\eea
Thus, the general integral is,
in the magnetic case and for $q\neq1$,
\be
\int\frac{dp}{\rho+p}=-(W^{(f)}+W^{(em)}),
\ee
\bea\label{W_q_neq0}
W_r=\mathcal{G}(r,\theta)+\ln(V_{sc})+g_{r}(\theta),
\\\nonumber
W_{\theta}=\mathcal{G}(r,\theta)+\ln(V_{sc})+g_{\theta}(r).
\eea

\section{The Boyer surfaces}\label{Sec:BoyerSurface}
The procedure described in the present article
borrows from the  Boyer theory on the equipressure surfaces for a PD torus \cite{Boy:1965:PCPS:,Pugliese:2012ub},  we consider  the equation for the \emph{Boyer potential}:
\(
W\equiv \mathcal{G}(r,\theta)+\ln(V_{sc})=\rm{c}=\rm{constant}
\),
where
 $\partial_{\theta} g_{r}(\theta)=\partial_{r} g_{\theta}(r)$  and  setting $g_{\theta}=g_{r}=0$,
 we can write
\be\label{E:2}
V_{sc}={\rm{K}} e^{-\mathcal{G}},
\ee
where ${\rm{K}}\equiv e^{\rm{c}}$, the toroidal surfaces  are obtained from the equipotential surfaces \cite{Boy:1965:PCPS:, Pugliese:2012ub}.
 eq.\il(\ref{E:2}) can be solved  for the radius $r$ assuming $q\approx0$.
We notice that, in the limit case $q=0$,
the magnetic field $B$, does not depend on the fluid enthalpy, furthermore  eq.\il(\ref{E:2}) in this case  is $V_{sc}={\rm{K}}$, this means that the magnetic field $B^{\phi}\left|_{q=0}\right.$ does not affect in anyway the Boyer potential and therefore the Boyer surfaces.

\begin{figure*}
\centering
\includegraphics[width=0.3\hsize,clip]{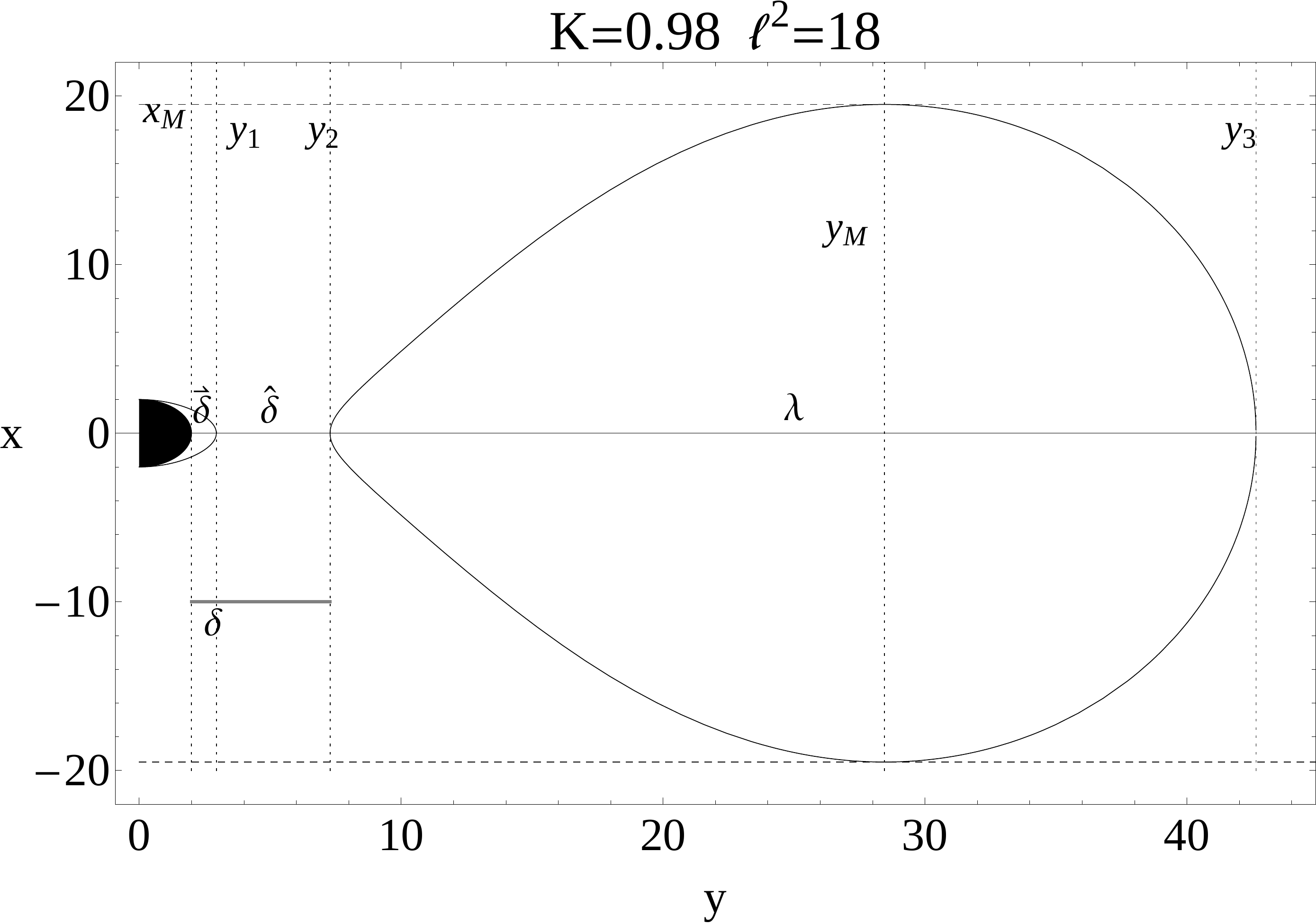}
\includegraphics[width=0.3\hsize,clip]{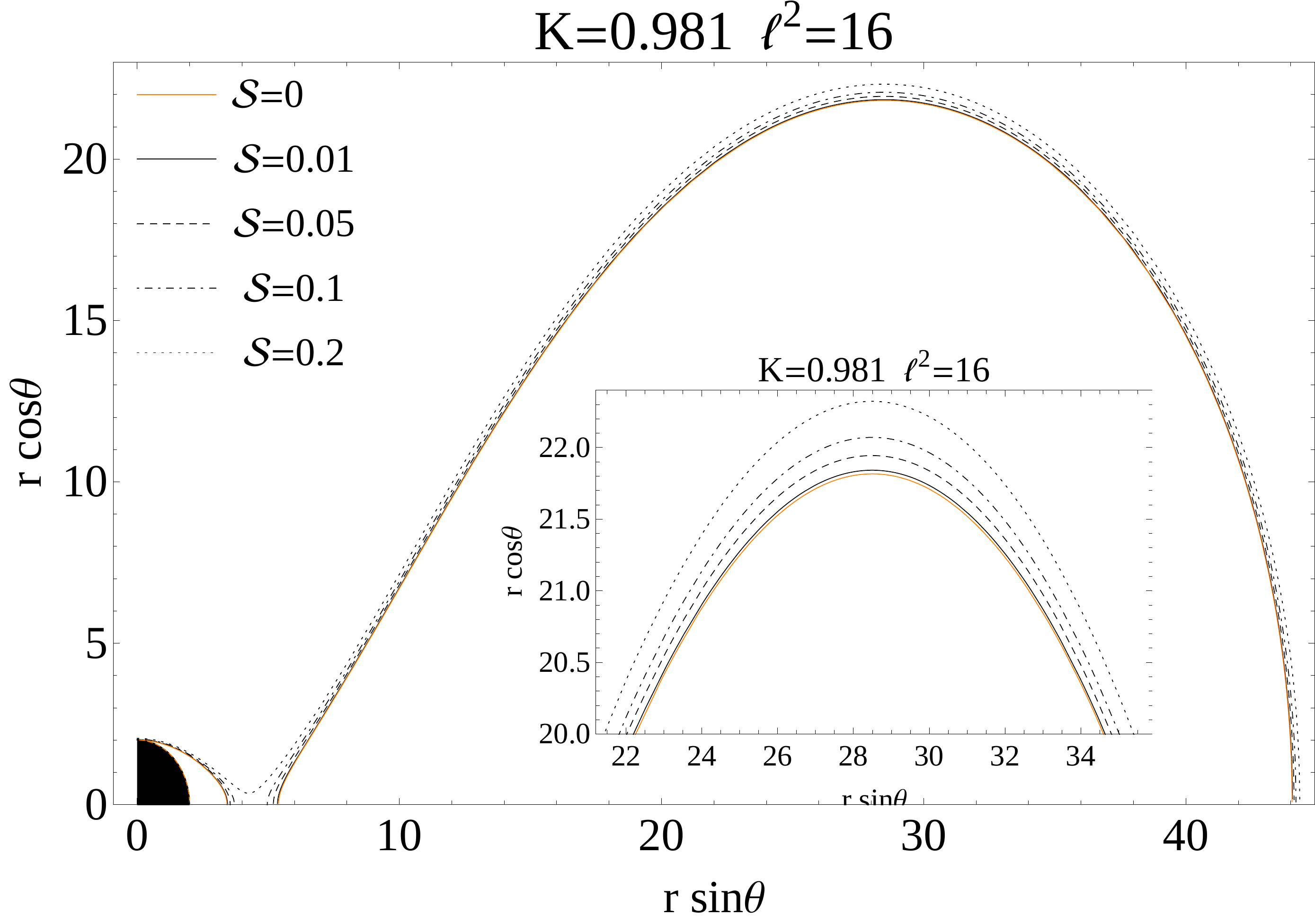}
\caption[font={footnotesize,it}]{(Colour on-line){Left} Case $B^{\phi}=0$:   the closed Boyer surface at ${\rm{K}}=0.981$ and $\ell^2=18$: $\lambda\equiv
y_3-y_2$ is the surface maximum diameter, $\delta\equiv y_2-2$ is the distance from the
source, $\hat{\delta}\equiv y_2-y_1$ is the distance from the inner surface,
$\breve{\delta}\equiv y_1-2$ is the distance of the inner surface from the horizon,
$h\equiv 2 x_M$ is the maximum thickness of the torus (torus height). {Right} Case $B^{\phi}\neq0$: torus surface at ${\rm{K}}=0.981$ and $\ell^2=16$,  in the plane $(r \sin \theta, r\cos\theta)$ for different values of $\Sie$. Inset figures are enlarged view of the main plot.}
\label{Griegn}
\end{figure*}

Consider now the case of small $q$,
eq.\il(\ref{E:2})  is
\be\label{E:kpot}
{\rm{K}} =1+\frac{\Sie }{ \sigma^2r(r-2)},
\ee
assuming the enthalpy $\omega$ to be a constant,
 we introduce the following $x=r \cos\theta$, $y=r \sin\theta$, and $\Sie=\Mie q/\omega$.
 In the following we consider  the approximation $\Sie\sim0$,
we note that the ratio  $\Mie/\omega$  gives the comparison between the magnetic contribution to the fluid dynamics, through $ \Mie $, and  the hydrodynamic   contribution   through its specific enthalpy $\omega$.
 {Solving eq.\il(\ref{E:kpot}), written in coordinates $(x,y)$, for  $x$ we obtain the  Boyer surfaces:}
\bea\label{Eq:xTx}
x^2=
\frac{4 \left({\rm{K}}^2 \ell^2+y^2\right)^2}{\left(y^2+{\rm{K}}^2 \left(\ell^2-y^2\right)\right)^2}-y^2+\frac{16 S \left({\rm{K}}^2 \ell^2+y^2\right)^2}{\left(y^2+{\rm{K}}^2 \left(\ell^2-y^2\right)\right)^3}.
\eea
The coordinates of the maximum point of the   surface satisfy the condition $dx/dy=0$ and $d^2x/d^2y<0$, this equation can be solved in order to obtain the exact form of $(x_{max},y_{max})$,
the torus thickness is here defined as the maximum height of the surface i.e. $h=2x_{max}$ with coordinate $y_{max}$.
We then define  the ratio
\be
R_s\equiv\frac{h}{\lambda},
\ee
as the \emph{squeezing function}  for the torus, where $\lambda=y_3-y_2$ is the maximum diameter of the torus surface section, see fig.\il\ref{Griegn}, and {$(y_1,y_2, y_3)$}, are  solutions of the equations  {$x^2=0$}.

Consider $R_s$, for  $\Sie\ll 1$:
\be\label{alice7}
R_s\approx R_s\left|_{\ti{($\Sie$=0)}}\right.+\frac{dR_s}{dS}\left|_{\ti{($\Sie$=0)}}\right. \Sie,
\ee
where $R_s(\Sie=0)$ is the  squeezing in the PD case $(\Sie=0)$, one can see this term  as the hydrodynamic contribution to the disk squeezing. Where the approximation $\Sie\ll 1$  holds we can study the linear dependence of $R_s$ from $\Sie$. The case of increasing (decreasing ) squeezing with $\Sie$ is defined by $\left(\frac{dR_s}{dS}\left|_{\ti{($\Sie$=0)}}\right.\right)>0\quad (\frac{dR_s}{dS}\left|_{\ti{($\Sie$=0)}}\right.<0)$. The Boyer surfaces  for  $\Sie=0$ and $\Sie\neq0$ are shown in figs.\il\ref{Griegn}.

\textbf{Remark:}
The magnetic pressure  is regarded  as a   perturbation of  the hydrodynamic component, it is assumed that the  Boyer theory remains   valid and applicable in this approximation (discussions on similar assumptions can be found in \cite{Komissarov:2006nz,
Abramowicz:2011xu}):  the torus  shape is determined by the  equipotential surfaces which now are regulated in eq.\il(\ref{W_q_neq0}) by an effective potential deformed, respect to  the PD case ($B=0$), by the magnetic field. The approximation $\Sie\approx0 $ follows from this assumption.
This analysis should be seen as a first attempt to compare the magnetic contribution to the torus dynamics  respect to the  kinetic pressure  in terms of the torus squeezing.
However,
we have performed several numerical analysis by setting a solution of the family of magnetic fields assigning a fixed value for the parameter $ q $.
The range of this parameter is clearly divided into two regions, say  $ 0 <q <1 $ and $ q> 1 $, with a subrange with extreme $ q = 2 $.
 The  case $q=2$ is indeed  interesting because the magnetic field loops  wrap around with toroidal topology along the torus surface.

For $ q <1$  the trend is clear: closed surfaces  are roughly those studied in the case $q\approx0$, for $ q> 1 $  we get  closed surfaces in the  limit $\Sie\approx0 $, which is in agreement with expansion around  $ \Sie=0 $, that is, when the contribution of the magnetic pressure to the torus dynamics is properly regarded as a perturbation  with respect to the hydrodynamics solution, this is realized when an appropriate condition $(\Sie\ll1)$  on the field parameters is satisfied.
%once $q$ is fi.
First we introduced the parameter $\Sie\equiv{\mathcal{M} q\omega ^{q-1}}/(q-1)$ in eq.\il(\ref{E:2}), then we expand  eq.\il(\ref{E:2})  around $ \Sie = 0$, and collect the terms up to first order. It should be noted that this equation, is equivalent to the corresponding  obtained  by an expansion in  $ q = 0 $, once  ignored the terms $  {\rm{K}}\Sie q$, and changing $ \Sie\rightarrow-\Sie$ (by convention we are taking $ \Sie> 0 $). The system analysed in eq.\il(\ref{Eq:xTx}) is thus  equivalent to the case of a torus with $\Sie=\left|\frac{\mathcal{M} q\omega ^{-1+q}}{-1+q}\right|\approx0$ and $(q {\rm{K}}\Sie)=O[\Sie]$  (i.e.   higher order than the 1th degree).
\section{Squeezing  of the Polish doughnuts}\label{Sec:SqueezingPD}
In this section we consider   the squeezing  of  the  PD model  $(B^{\phi}=0)$.
The PD torus has been extensively studied in the literature with both analytical and numerical methods,  we refer in particular to \cite{Pugliese:2012ub} with which we share much of the notation and conventions used in this work.
Setting $\Sie=0$  in eq.\il(\ref{Eq:xTx}) we find

\be\label{super}
x=\pm\sqrt{\left[\frac{2 \left({\rm{{\rm{K}}}}^2 \ell^2+y^2\right)}{{\rm{{\rm{K}}}}^2
\left(\ell^2-y^2\right)+y^2}\right]^2-y^2},
\ee
 with $2\sqrt{2}/3<{\rm{K}}<1$ and angular momentum
$({\ell}_k^-)^2<{\ell^2}<({\ell}_k^+)^2$, where ${\ell^2}_k^->27/2$,  we refer to \cite{Pugliese:2012ub} for the exact expression of  ${\ell}_k^{\pm}$   and details.
fig.\il\ref{Griegn}-right shows a    section  of the PD torus for  ${\rm{K}}=0.981$ and $\ell^2=18$.
We summarize the results concerning the squeezing function $ R_s = h / \lambda $  in the following.

\begin{figure*}
\centering
\includegraphics[width=0.3\hsize,clip]{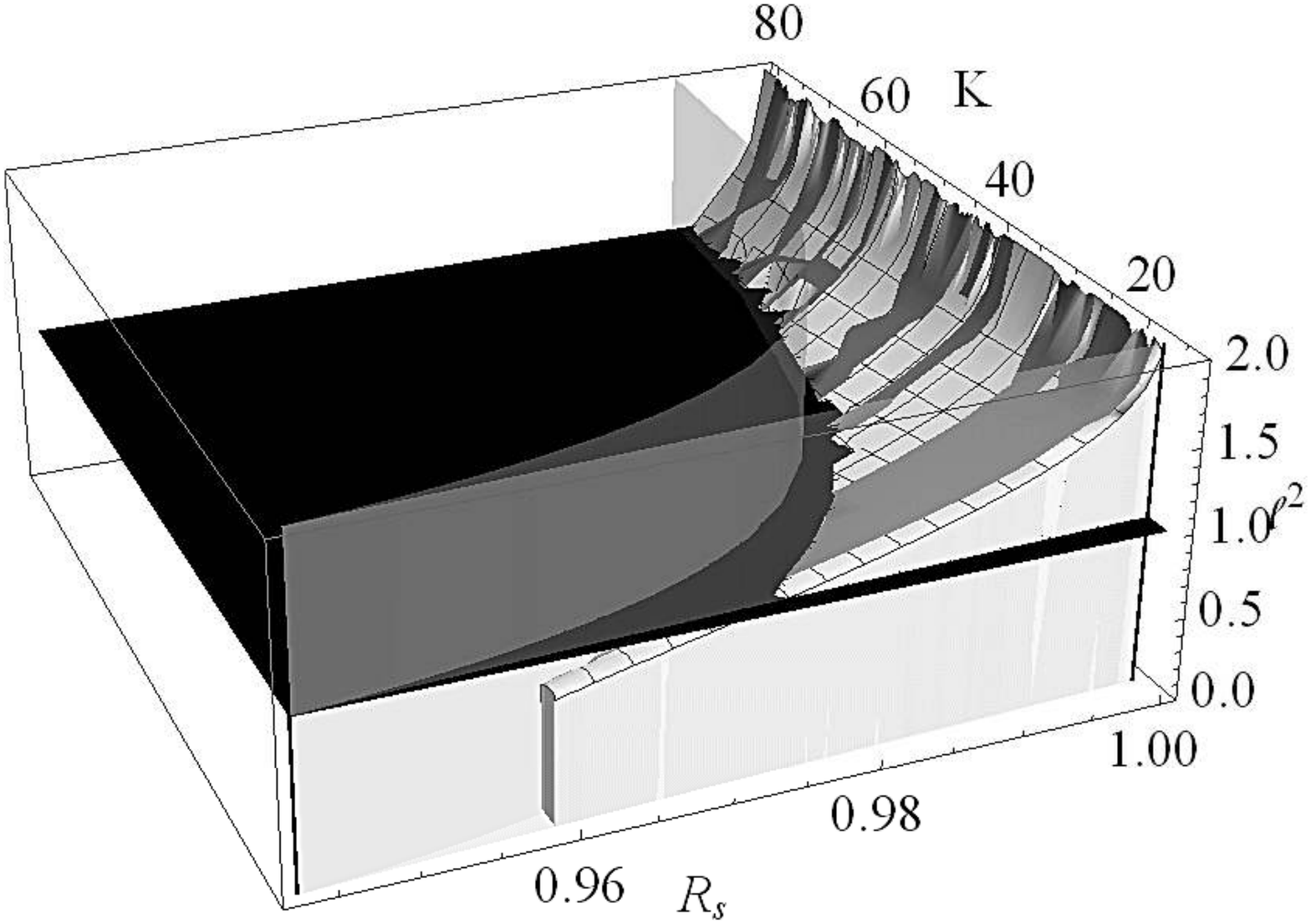}
\includegraphics[width=0.3\hsize,clip]{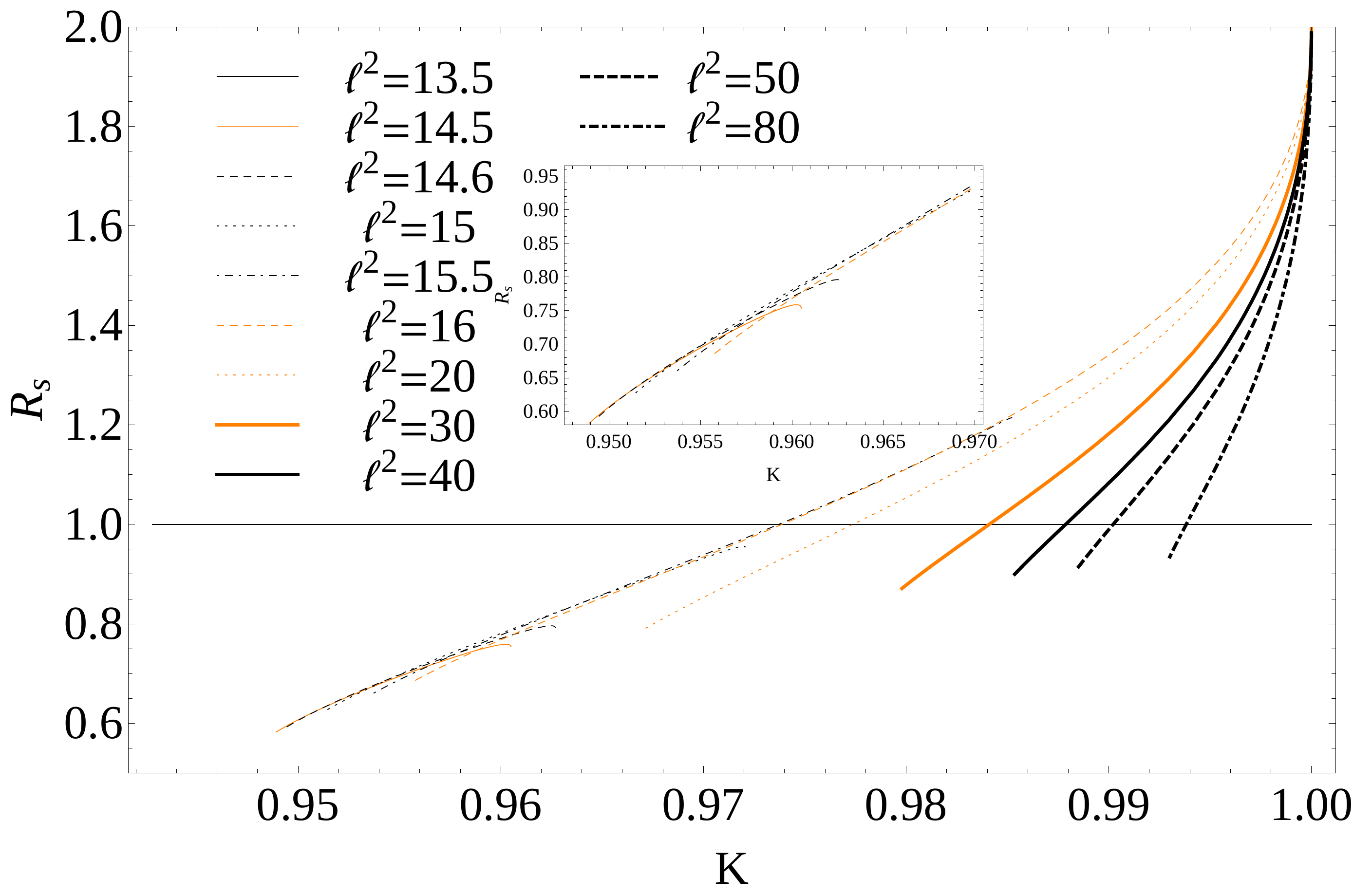}
\includegraphics[width=0.3\hsize,clip]{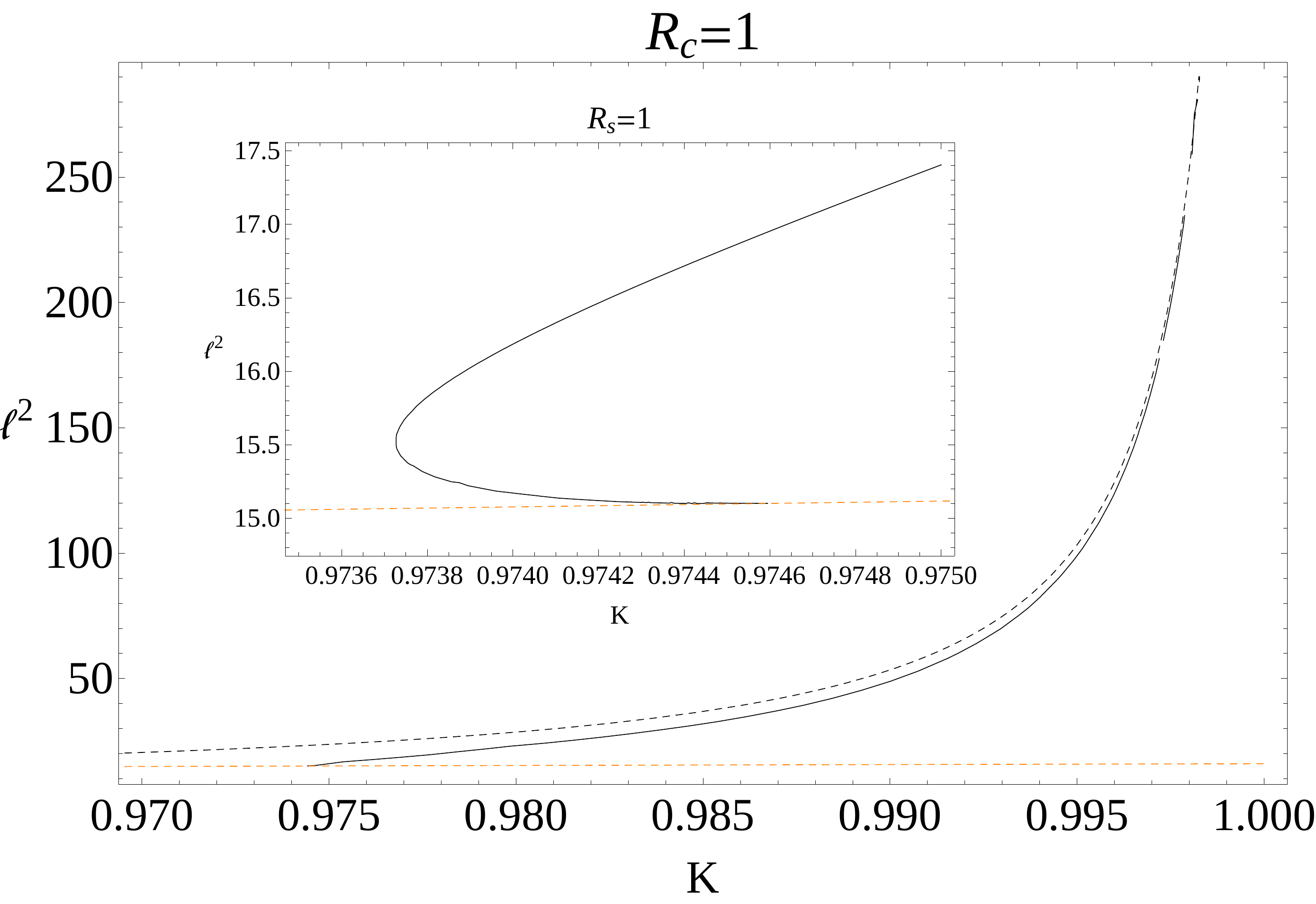}
\caption[font={footnotesize,it}]{\footnotesize{(Colour on-line) Case $B^{\phi}=0$. Left: The squeezing  $R_s$ as function of  ${\rm{K}}$ and the angular momentum   $\ell^2$. The black plane is $R_s=1$,  the curves $({\ell}_k^{\pm})^2$  are plotted. Center: $R_s$ curves for fixes values of the $\ell^2$ as function of ${\rm{K}}$, a zoom of the curves  is in the inset plot: $R_s$ increases monotonically with ${\rm{K}}$ and, at fixed ${\rm{K}}$, decreases with $\ell^2$.  Right: curve $R_s=1$ (black curve), orange dashed line  is $(\ell_k^-)^2$ black dashed line is $(\ell_k^+)^2$, inside plot is a zoom of the   $R_s=1$ in the region ${\rm{K}}\in[ 0.9735, 0.975]$}} \label{evarc}
\end{figure*}

\begin{figure*}
\centering
\includegraphics[width=0.3\hsize,clip]{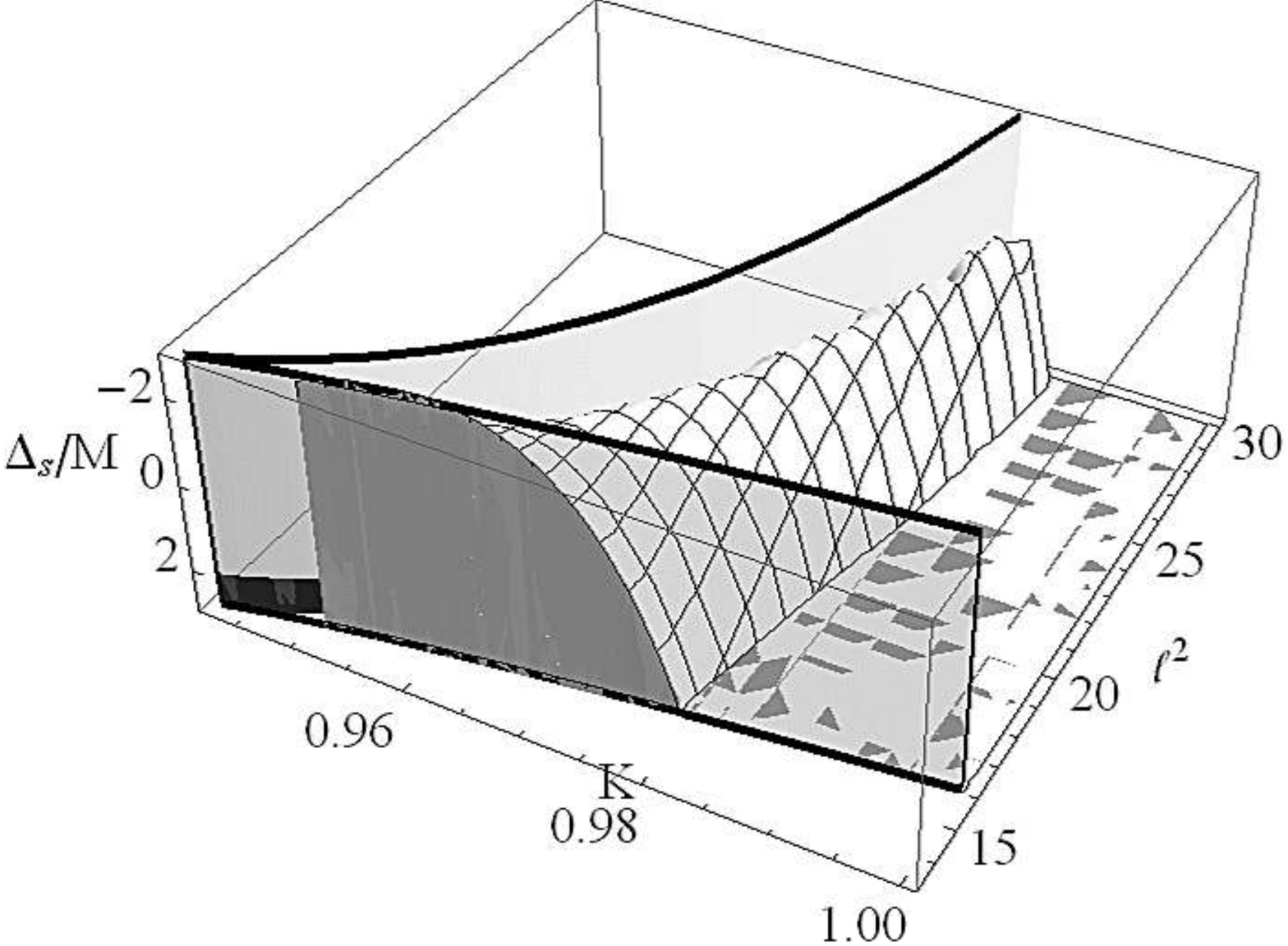}
\includegraphics[width=0.3\hsize,clip]{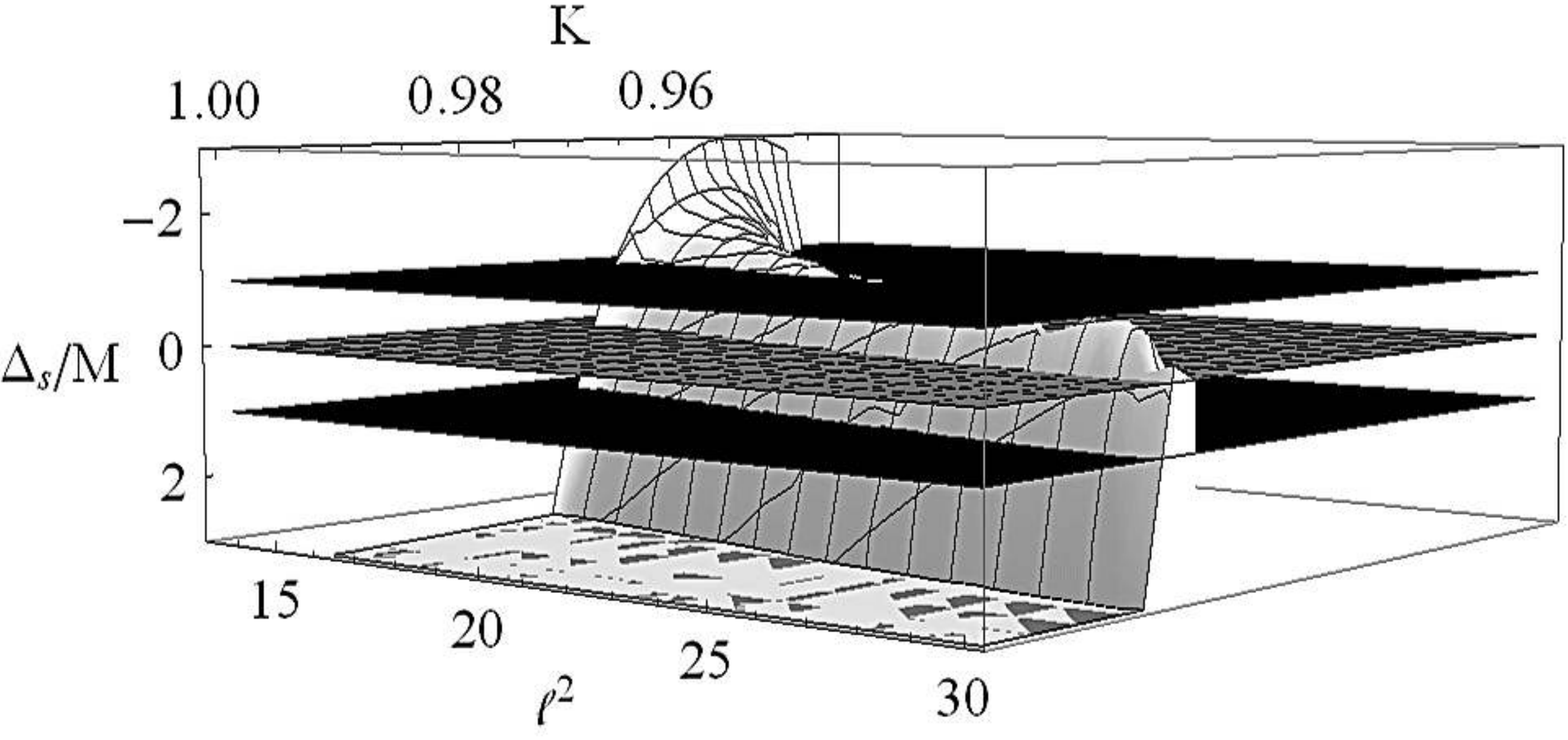}
\includegraphics[width=0.3\hsize,clip]{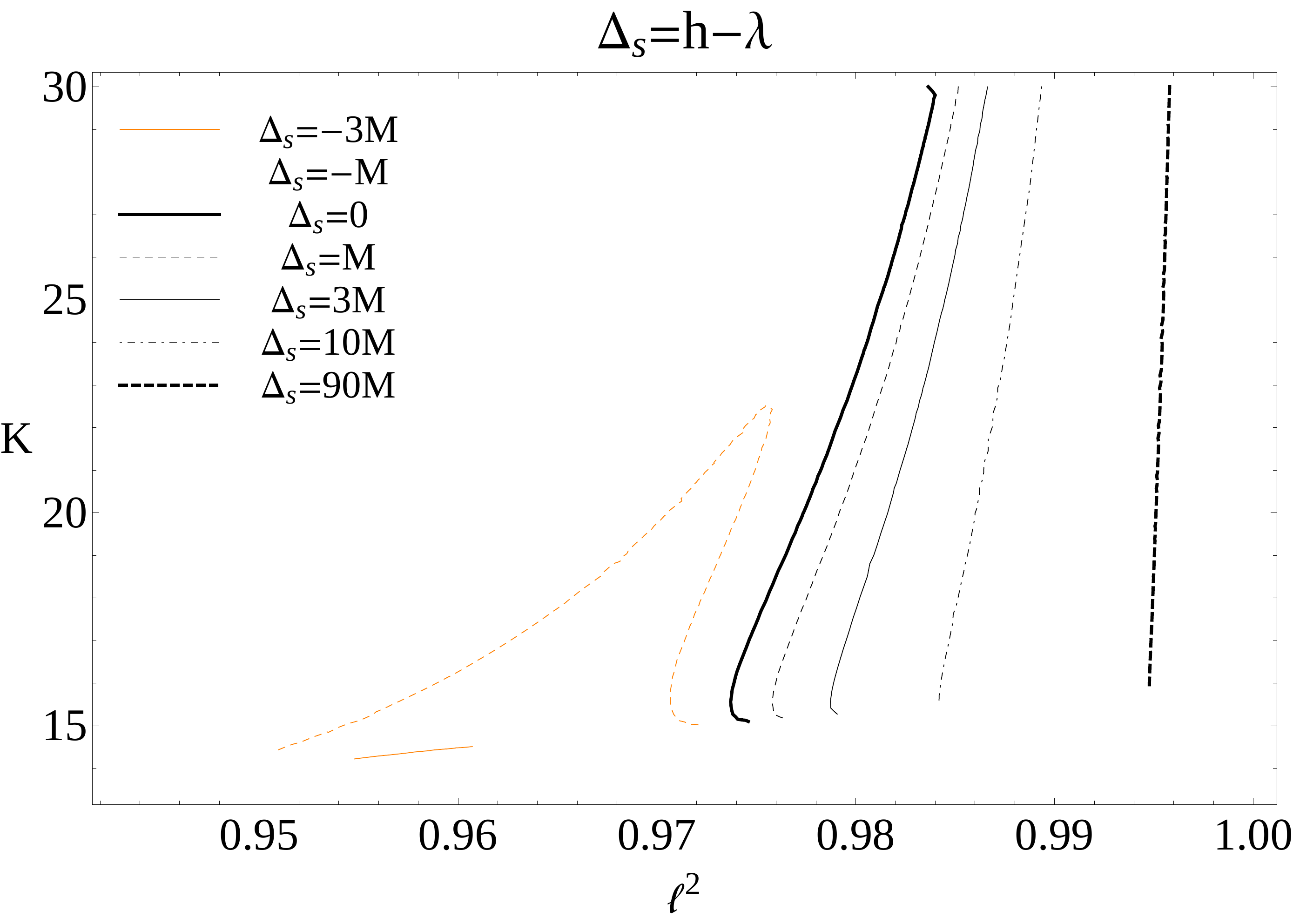}
\caption[font={footnotesize,it}]{\footnotesize{(Colour on-line) Case $B^{\phi}=0$. Left: difference $\Delta_s\equiv h-\lambda$ as function of ${\rm{K}}$ and $\ell^2$. Center plot shows $\Delta_s$ and the planes $\Delta_s=(0,\pm2)$. Left: curves  $\Delta_s=${\rm{cons}} as function of $\ell^2$ and ${\rm{K}}$.}} \label{volonta}
\end{figure*}

The squeezing function  $R_s$   increases monotonically with ${\rm{K}}$ and, at fixed ${\rm{K}}$ decreases with $\ell^2$, figs.\il\ref{evarc}.
This evolution can also be seen by studying of the function $\Delta_s\equiv h-\lambda$, shown in figs.\il\ref{volonta},\ref{full3dp} as function of $\ell$ and ${\rm{K}}$.
 However we note that the lower $R_s\equiv h/\lambda$  is and the thinner is the torus, conversely higher is $ R_s  $ and thicker is the torus.
This therefore means that the toroidal surface is squeezed on the equatorial plane with decreasing  $ {\rm{K}} $ and  increasing $\ell^2$.
These  results are indeed  consistent with the analysis in \cite{Pugliese:2012ub} where it  was shown  that
 the torus thickness  $h$, increases with the
 ${\rm{K}}$ and decreases with the angular momentum, that is the torus becomes thinner for
    high angular momenta, and  thicker  for high energies.
The { maximum diameter} $\lambda$ increases with the ${\rm{K}}$, but
decreases with the fluid angular momentum, moreover the location of maximum thickness of the torus moves  towards the external
regions  with increasing angular momentum and ${\rm{K}}$, until it reaches a maximum    an then decreases.

We note    a region  of   $\ell$ and ${\rm{K}}$ values, in which this trend is reversed, that is
$ R_s =h/\lambda<1 $ figs.\il\ref{evarc}- (left-center), and  $\Delta_s=h-\lambda<0$ figs.\il\ref{volonta},\ref{full3dp}-left, i.e. the squeezing is such that the disk is longer that thick, although the minimum value  $ R_s \approx0.95 $  is still quite large because it still makes sense that the model of PD be a thick disk. One sees  from  figs.\il\ref{full3dp},   in this region, a minimum for $\Delta_s$ as a function of $\ell$. Following the curve  for increasing values of ${\rm{K}}$, the function $\Delta_s$ decreases or decreases until it reaches  a minimum and then increases.
From fig.\il\ref{evarc}-\emph{center } one also sees that the lower the angular momentum is and more configurations with $ R_s <1 $ are.
Finally, some curves  overlap in this region for one or more values of ${\rm{K}}$, this  means that  torus configurations would exist  with different values of $ \ell^2 $, but equal  $ {\rm{K}} $ and $R_s$.

Fig.\il\ref{evarc}-right
shows the solutions of equation $R_s =1$, on the curve $ \ell^2({\rm{K}})$ .
Essentially this curve confirms the previous results showing that the torus where the $(h)$ equals the length on the equatorial plane $\lambda$ increases in angular momentum  with ${\rm{K}}$, first slowly and then more and more rapid as ${\rm{K}}$ approaches the value limit $ {\rm{K}} = 1$. A fixed $R_s $, the centrifugal effects must compensate for an increase of ${\rm{K}}$ which tends to bring the disk to a greater $R_s $, and to sufficiently small values of  $ {\rm{K}} \in(0.97367 ,0.9746)$, the curve bends back on itself, {see Figs.\il\ref{evarc}}.
\section{Squeezing of the  magnetized torus}\label{Sec:SqueezingMT}
In this section we summarize  the
 analysis concerning the squeezing of the magnetized torus $(\Sie\neq0)$.

\begin{figure*}
\centering
\includegraphics[width=0.3\hsize,clip]{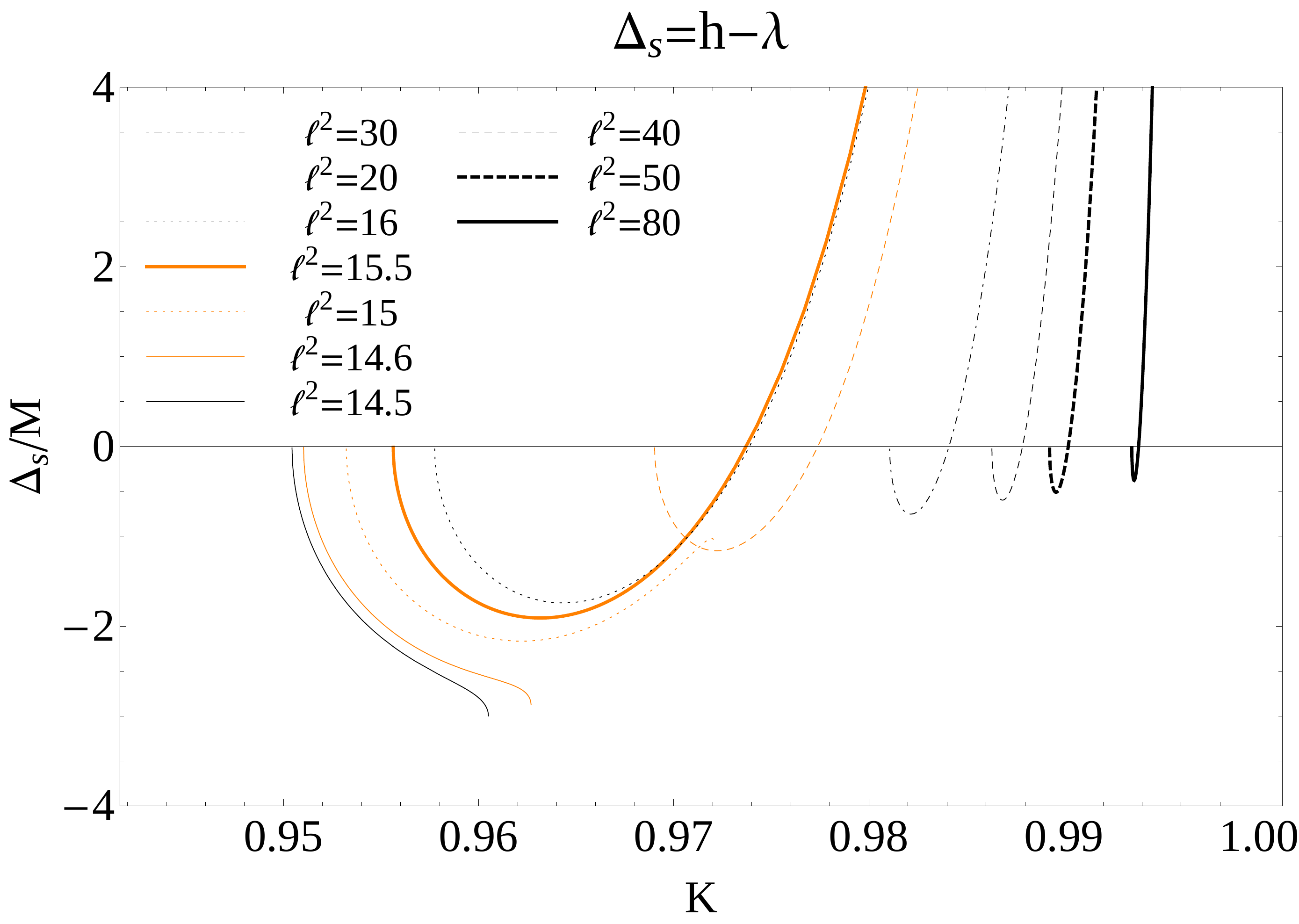}
\includegraphics[width=0.3\hsize,clip]{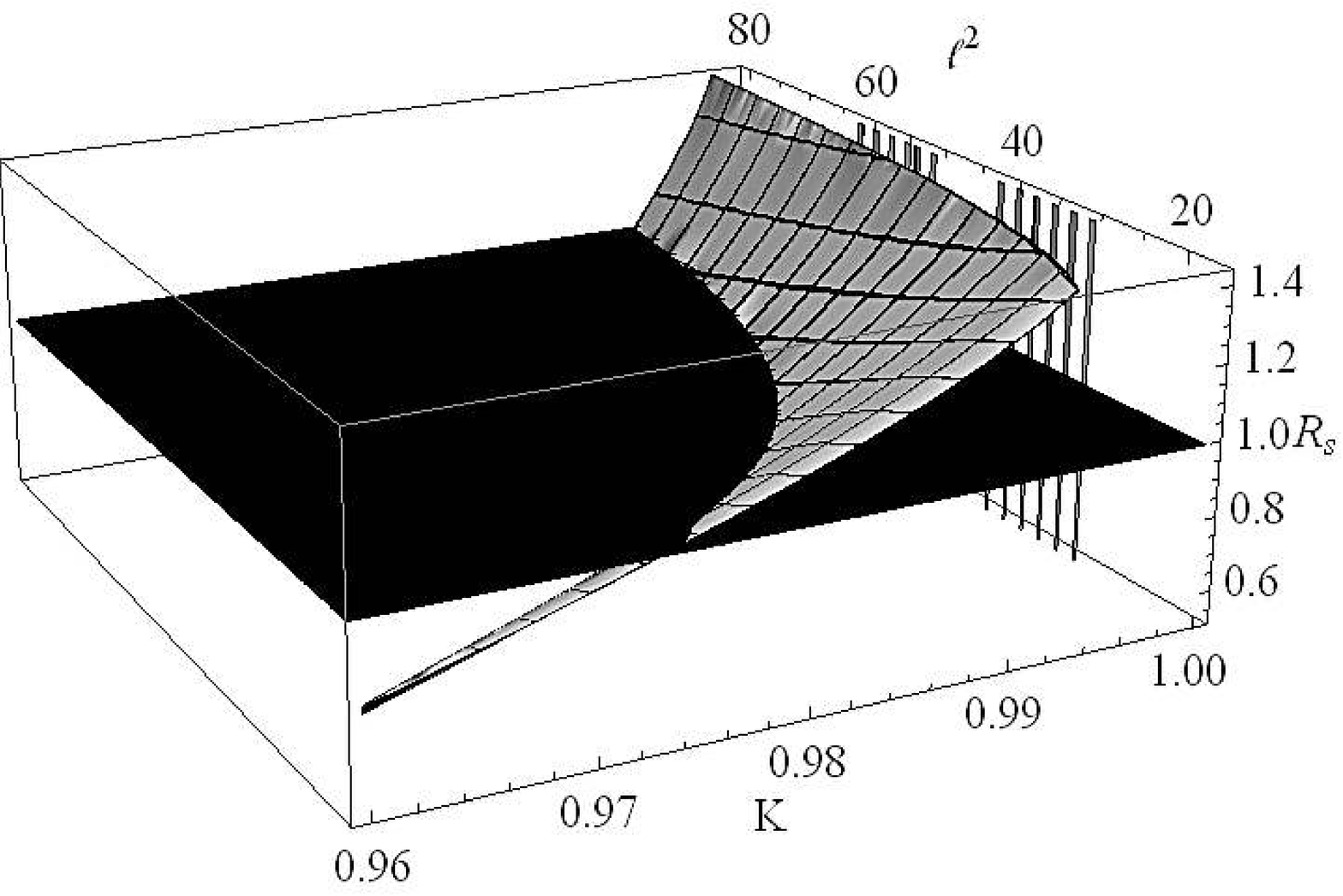}
\includegraphics[width=0.3\hsize,clip]{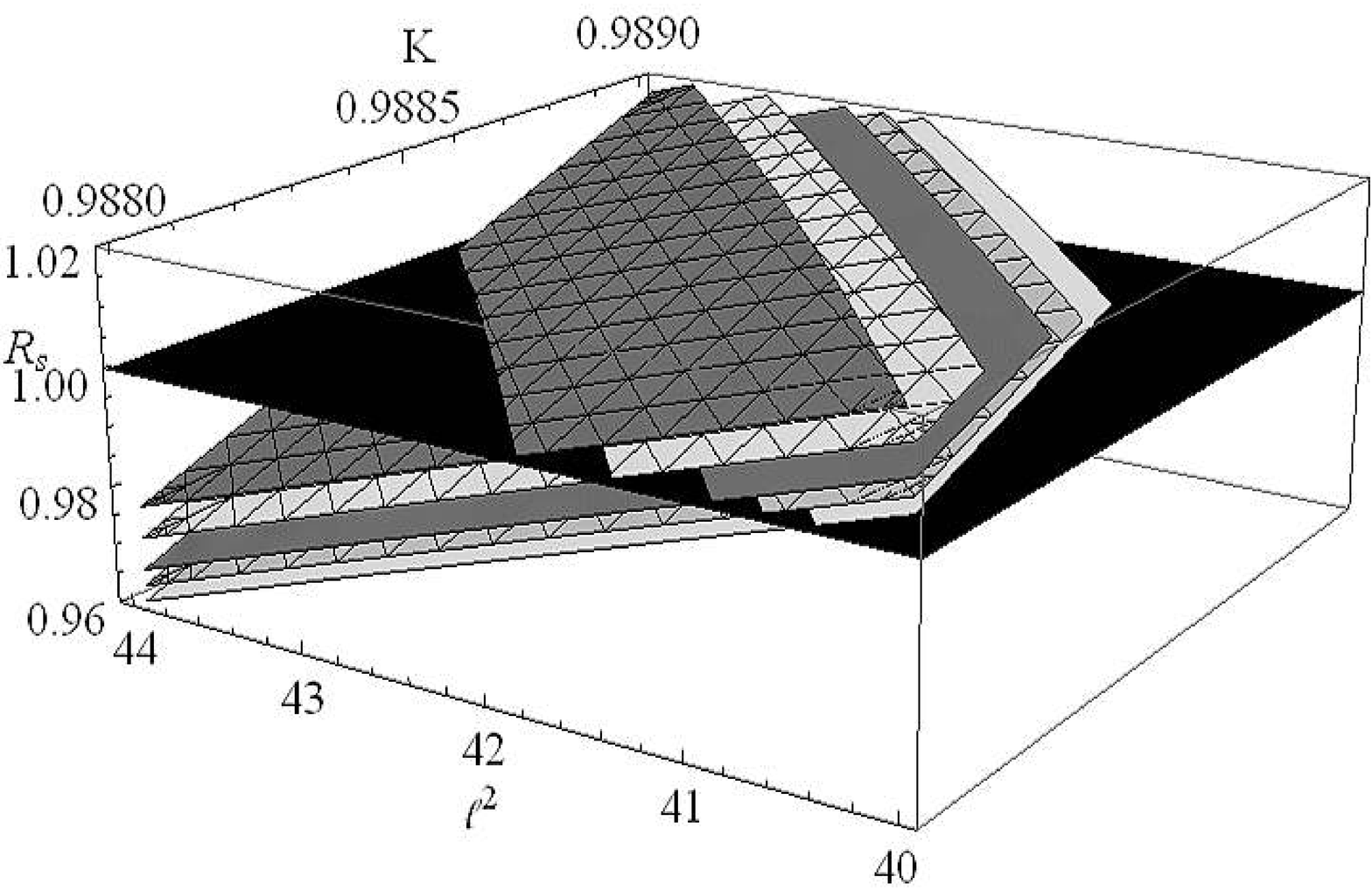}
\caption[font={footnotesize,it}]{\footnotesize{(Colour on-line) Left: Case $B^{\phi}=0$, plot of $\Delta_s$ as function of ${\rm{K}}$, for selected values of $\ell$.  Case $B^{\phi}\neq0$. Center: the squeezing $R_s$ as function of  ${\rm{K}}$ and the angular momentum with $\ell^2$, for selected values of the $\Sie$ parameter. The black plane is $R_s=1$. A zoom is shown in right plot: white surface $\Sie=0$, shaded lightgray surface $\Sie=1/8$, gray surface $\Sie=1/4$, shaded white surface $\Sie=1/2$, shaded gray surface $\Sie=3/4$.} }\label{full3dp}
\end{figure*}

\begin{figure*}
\centering
\includegraphics[width=0.3\hsize,clip]{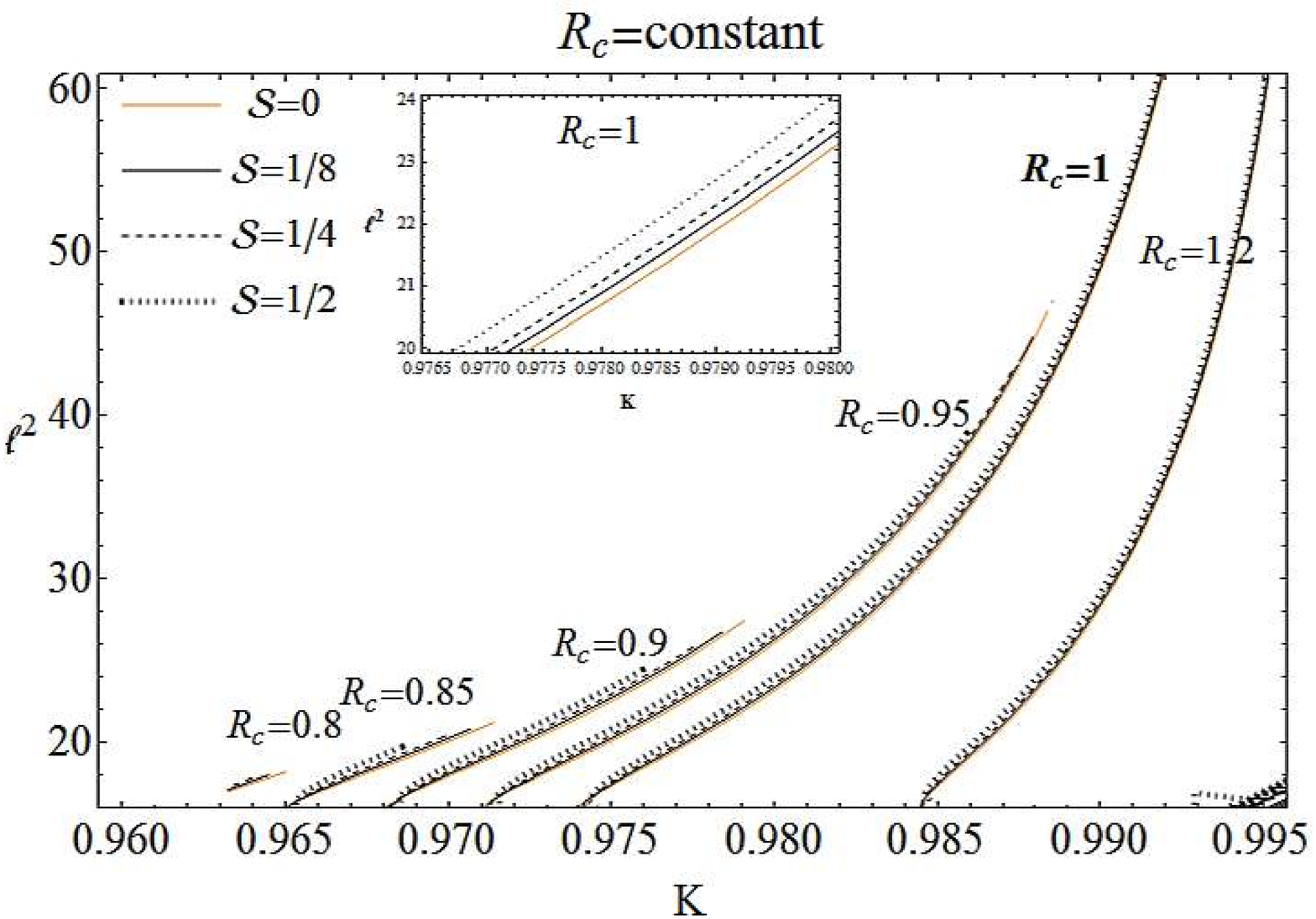}
\includegraphics[width=0.3\hsize,clip]{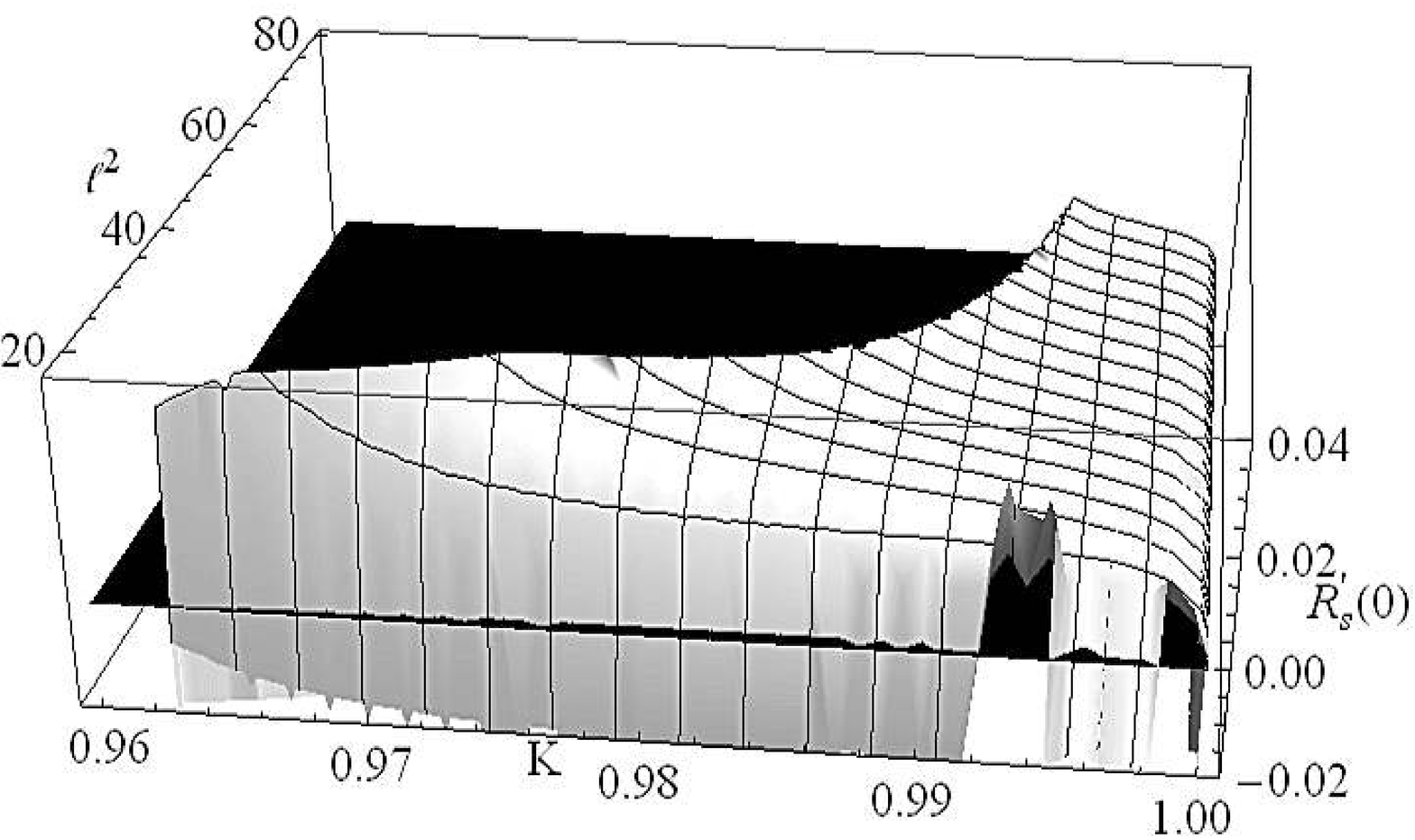}
\includegraphics[width=0.3\hsize,clip]{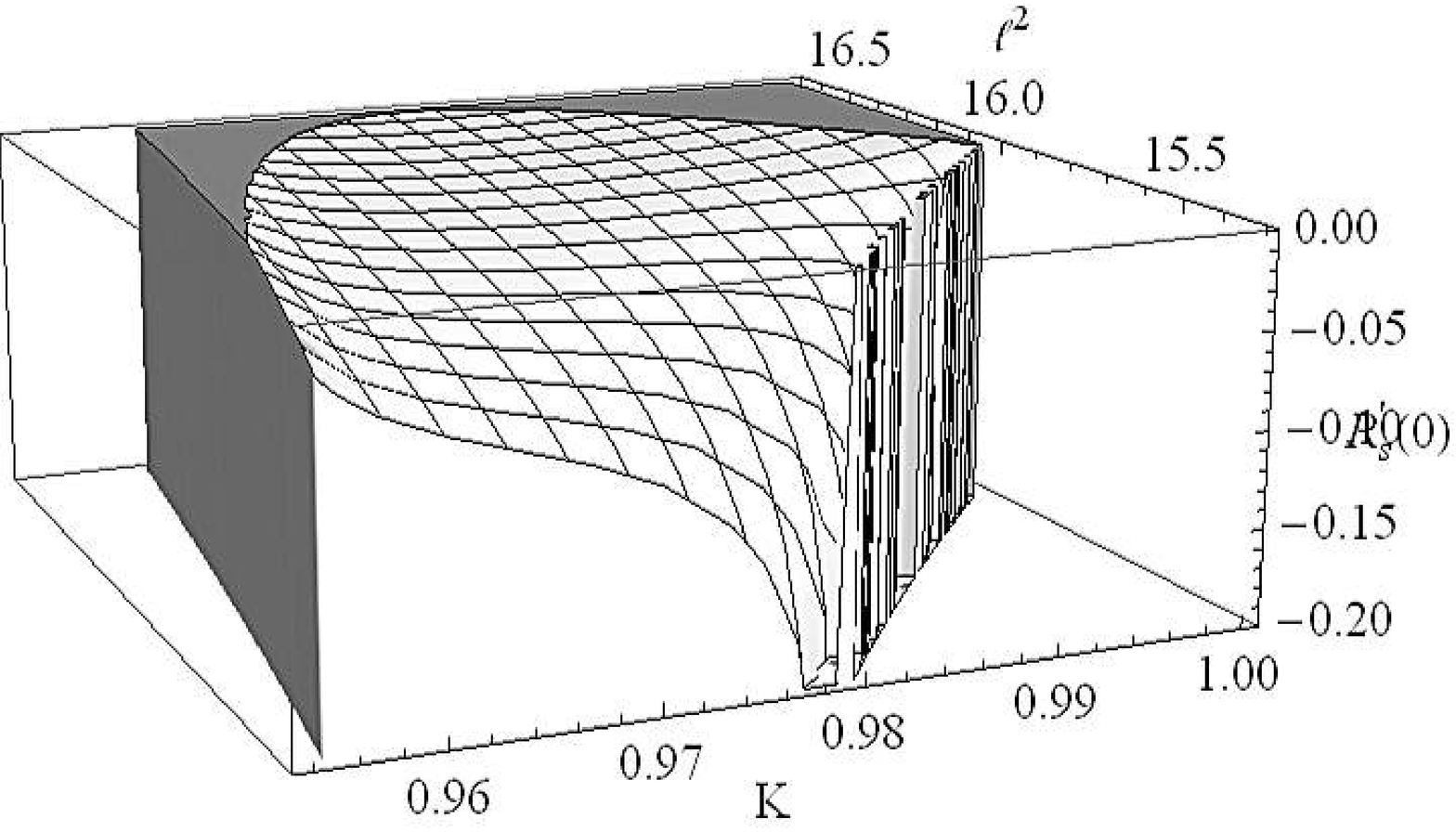}
\caption[font={footnotesize,it}]{\footnotesize{Case $B^{\phi}\neq0$.  Left: the squeezing
surface $R_s=$cons  for selected values of $\Sie$ as function of $\ell^2$ and ${\rm{K}}$. The inset plot shows the curves $R_s=1$ in the region ${\rm{K}}\in(.9765,.98)$ $\ell^2\in(20,24)$. Center: the function $R'_c(0)\equiv dR_s/d\Sie$ in $\Sie=0$ as function of the $\ell^2$ and the parameter ${\rm{K}}$. The function is potted in the range $R'_c(0)\in(-0.02,+0.02)$, black plane is $R'_c(0)=0$.
Right plot is a zoom in the region $R'_c(0)\in(-0.02,0)0$, ${\rm{K}}\in(0.954, 1)$, $\ell^2\in(15.3,16.65)$.
} }\label{RPartita}
\end{figure*}

Fig.\il\ref{full3dp} shows several surfaces $R_s (\ell^2, {\rm{K}})$  for different values of $\Sie$, an enlarged view of this plot,
distinguishes the different  $\Sie$-surfaces.
In general the function $R_s (\ell^2, {\rm{K}})$,  as for  the case $\Sie = 0$,
increases with ${\rm{K}}$ but decreases with $\ell^2$ that is  the torus is thicker  as the ${\rm{K}}$ parameter increases and becomes thinner as fluid angular momentum increases, fig.\il\ref{full3dp}-center, furthermore $R_s$ increases with $\Sie$, fig.\il\ref{full3dp},\ref{RPartita}-left.
%\\
%\item[]

\begin{figure*}
\centering
\includegraphics[width=0.3\hsize,clip]{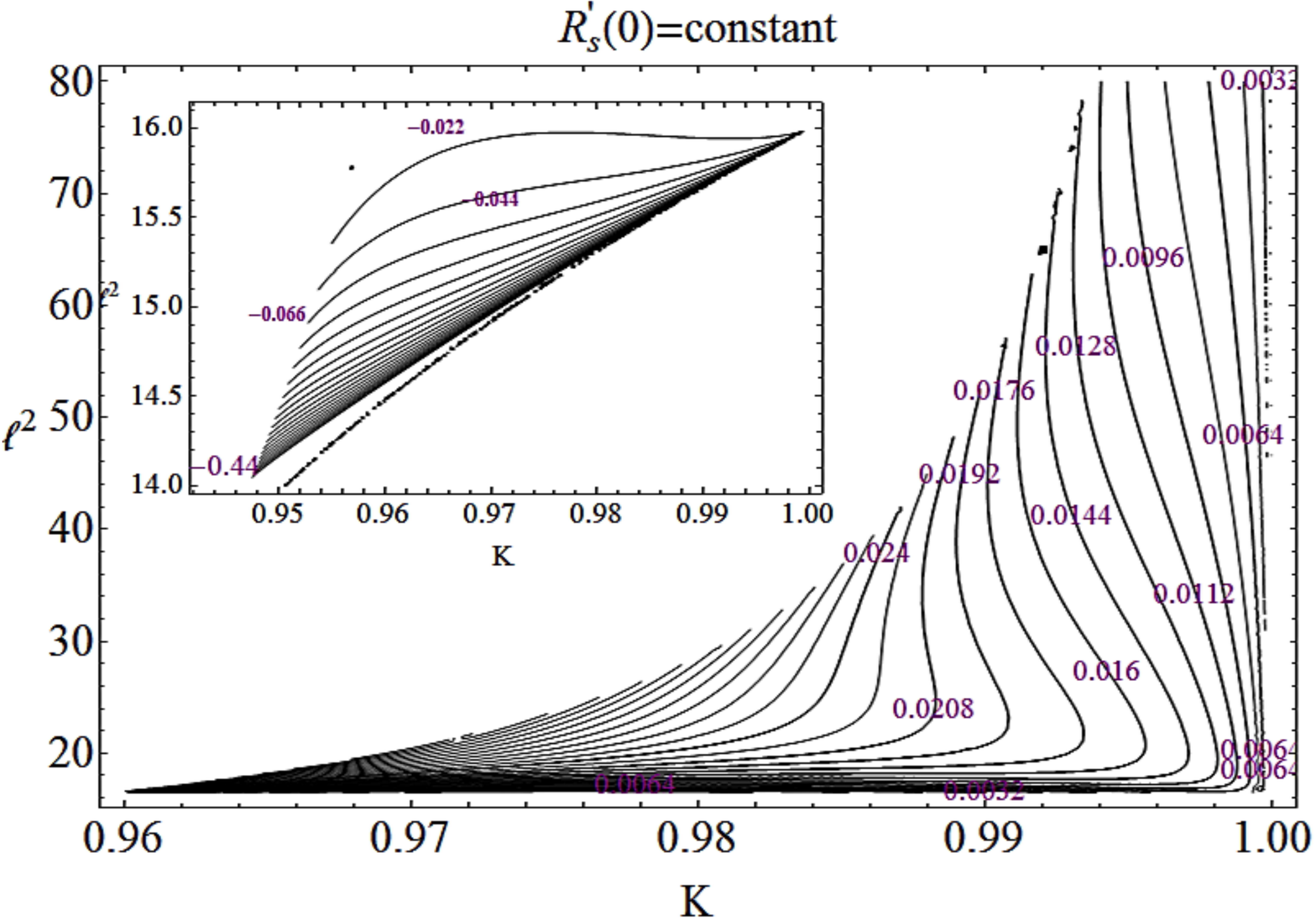}
\includegraphics[width=0.3\hsize,clip]{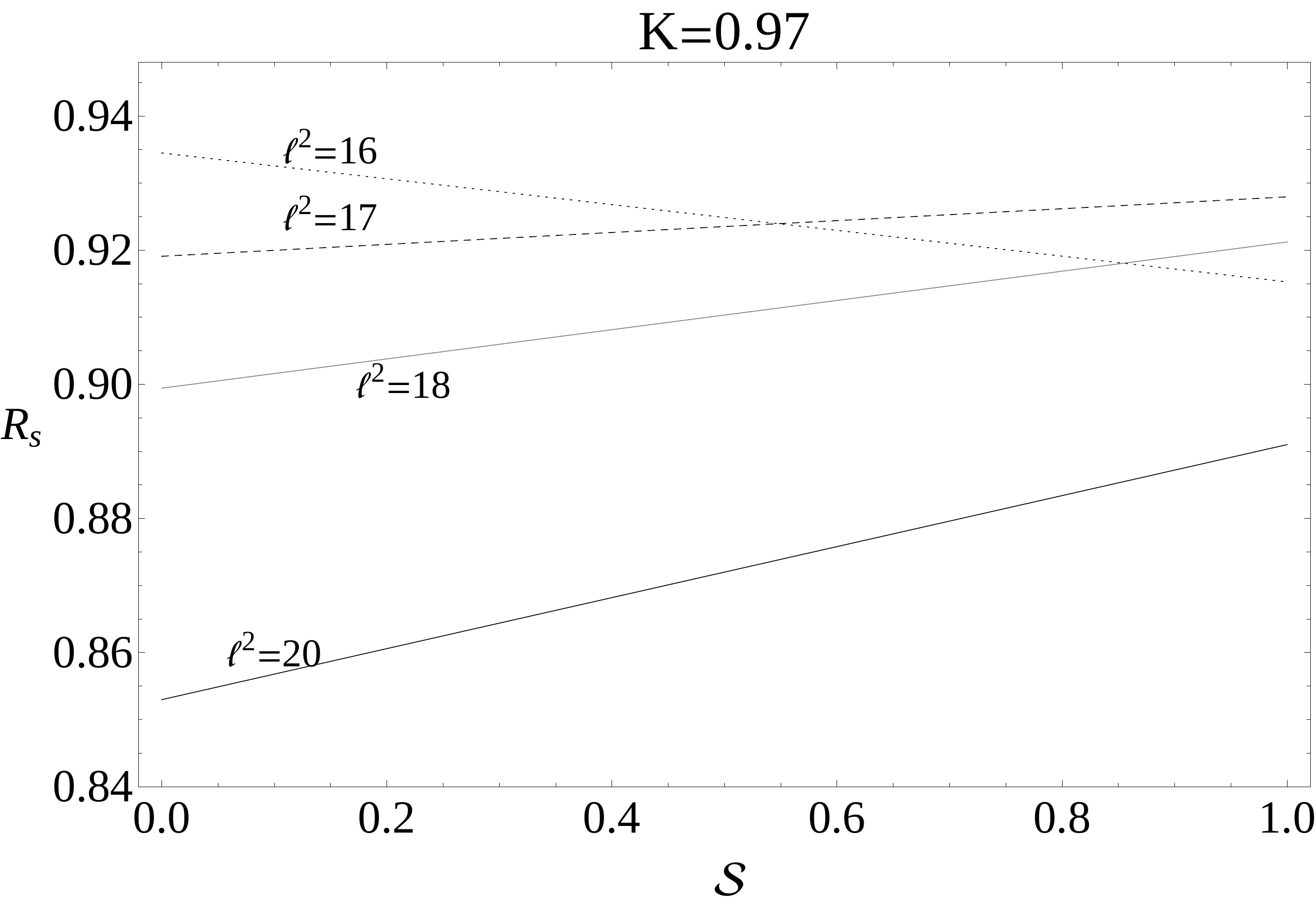}
\includegraphics[width=0.3\hsize,clip]{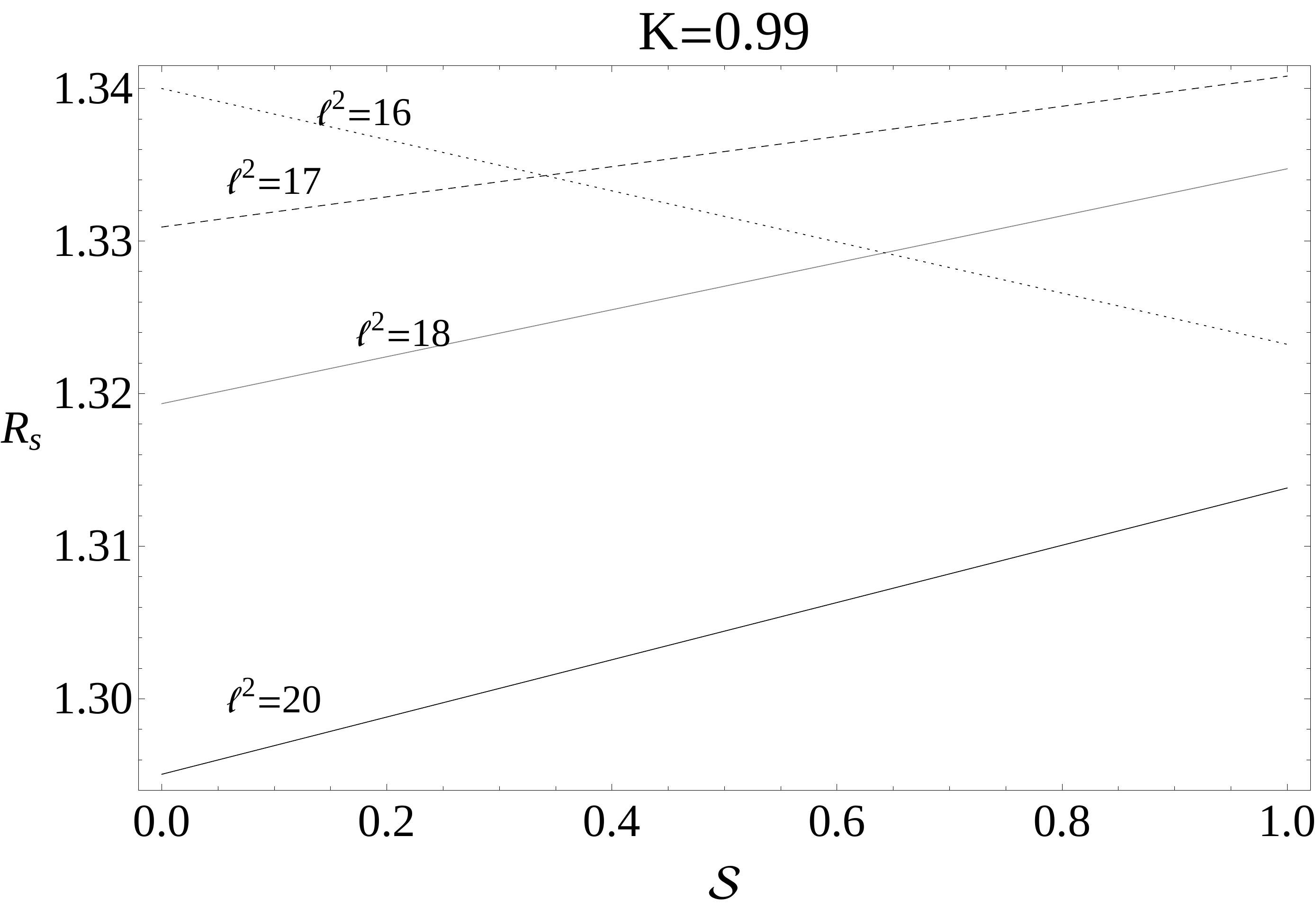}
\caption[font={footnotesize,it}]{\footnotesize{Case $B^{\phi}\neq0$. Left: curves $R'_c(0)=$constant in the plane $(\ell^2,{\rm{K}})$, inside plot is a zoom in the region  $\ell^2\in(15.3,16)$ where $R'_c(0)<0$. Plot of the squeezing $R_s$ as function of $\Sie$ for selected values of $\ell^2$ and  ${\rm{K}}=0.97$ (center plot), ${\rm{K}}=0.99$ (right plot).  Plots show the range $\Sie\in[0,1]$, however the approximation $\Sie<<1$ holds for $\Sie<0.3$.} }\label{SPsVilia}
\end{figure*}

The derivative $(dR_s/d\Sie) (\Sie=0)$ is overwhelmingly positive: for the magnetized torus,
$R_s $  grows with $\Sie$, hence the torus is thicker as $\Sie$ increases and thinner as $\Sie$ decreases, figs.\il\ref{RPartita},\ref{SPsVilia}-left.
However, {as evidenced by the Figs,\il\ref{RPartita},\ref{SPsVilia}} there is a small region  $15.3\leq\ell^2\lesssim 16.65$, in which the derivative $(dR_s/d\Sie) (\Sie=0)$  is negative and $R_s$ is a decreasing function of  $\Sie$:  from this we conclude that
$R_s$ decreases compared to the case $\Sie = 0$ i.e. the torus becomes thinner with increasing of $\Sie$ and viceversa becomes  thicker to the decrease of $\Sie$, figs.\il\ref{RPartita}.
These results  are confirmed   in fig.\il\ref{SPsVilia}, where the  squeezing  has been plotted as a function of  $\Sie$ and $\ell^2$, for different values of ${\rm{K}}$, however it should be noted that although the figures show the behavior of the ratio $R_s$ across the range $0<\Sie<1$, the approximations assumed in this work are valuable for small $\Sie$. Fig.\il\ref{SPsVilia} shows that $ R_s$ is generally increasing with ${\rm{K}}$ and decreasing with $\ell^2$, and also shows the existence of a region where this trend is reversed.
\section{Conclusions}\label{Sec:Discussion}
We have studied the Polish doughnut (PD) model   of thick disk  for a   perfect fluid, circularly orbiting around a Black hole of Schwarzschild, in an  hydrodynamic and magnetohydrodynamic context, using the effective potential  approach examined in detail for the  hydrodynamical model in \cite{Pugliese:2012ub}. For the magnetohydrodynamic model we used the solutions of the barotropic torus with a toroidal magnetic field  discussed in \cite{Komissarov:2006nz,Montero:2007tc}. We have considered the magnetic contribution to the total pressure of the fluid as a perturbation of the effective potential for the exact full general relativistic part.
Once defined the torus squeezing function $ R_s \equiv h / \lambda $   as the ratio maximum torus height-to-maximum diameter  of the torus  section,
we studied $R_s$  varying of the angular momentum $ \ell $, the effective potential $ {\rm{K}} $ and the parameter $ \Sie $, the last evaluates the magnetic contribution to the fluid dynamics respect to   the hydrodynamic   contribution   through its specific enthalpy.
The  lower is $R_s$   and the thinner is the torus, conversely the higher is $ R_s$ and thicker is the torus.
It is shown that  given the presence of a magnetic pressure as a perturbation of
the hydrodynamic component, there are not quantitatively significant effects on the thick disk model; however the analysis showed that the squeezing function  has in general a monotonic trend with $(\ell, {\rm{K}}, \Sie)$. For the hydrodynamic case
the squeezing function  $R_s$   increases monotonically with ${\rm{K}}$, and at fixed ${\rm{K}}$ decreases with $\ell^2$,
this therefore means that the toroidal surface is squeezed on the equatorial plane with decreasing ``energy'' ${\rm{K}}$ and  increasing $\ell^2$, these results are indeed in agreement with analysis in \cite{Pugliese:2012ub}.
However we observed     a region  of   $\ell$ and ${\rm{K}}$ values, in which this trend is reversed, that is
$R_s=h/\lambda<1$ reaching a minimum values of  $R_s\approx0.95$,
for increasing values of ${\rm{K}}$, the squeezing  decreases or decreases until it reaches   a minimum and then increases.
In the  magnetohydrodynamic case, we evaluated the influence of the magnetic field through the parameter  $\Sie$.
In general, as for  the case $\Sie=0$,  the torus is thicker  as the ${\rm{K}}$  parameter increases, and becomes thinner as fluid angular momentum increases, furthermore $R_s$ increases with $\Sie$.
However there is a  small region of $\ell$, in which
$R_s$ is a decreasing function of  $\Sie$:  that is
$R_s$ decreases compared to the case $\Sie = 0$ i.e. the torus becomes thinner with increasing of $\Sie$ and viceversa becomes  thicker with decreasing $\Sie$.

In conclusion, this analysis shows that the torus thickness  is not  modified in a  quantitatively significant way by the presence of a toroidal magnetic field although it is much affected by the variation of the  angular momentum.
However we consider  this analysis as a first attempt to compare the magnetic contribution to the torus dynamics with respect to the kinetic pressure  in terms of  the torus squeezing and, more in general, the plasma confinement  in the disk toroidal surface.
\acknowledgments
This work has been developed in the framework of the CGW Collaboration
(www.cgwcollaboration.it). DP gratefully acknowledges financial support from the Angelo
Della Riccia Foundation and  wishes to thank the Blanceflor Boncompagni-Ludovisi, n\'ee Bildt 2012.


\begin{thebibliography}{0}
\bibitem{FeB04}
 Fender R., Belloni T., 2004,
 { Ann. Rev. Astron. Astrophys.}, {42}, 317.

 \bibitem{So07}
Soria R., 2007,
\newblock {Astrophysics and Space Science}, 311, 213.

 \bibitem{Balbus2011}Balbus S.A.,  2011, Nat., 470, 475

\bibitem{Shakura1973} Shakura N.I.,  Sunyaev R.A.,
 1973, A\&A,  24, 337-355.

 %\cite{Abramowicz:2011xu}
\bibitem{Abramowicz:2011xu}
Abramowicz M.~A.~ and Fragile P.~C.~,
  %``Foundations of Black Hole Accretion Disk Theory,''
  arXiv:1104.5499.
  %%CITATION = ARXIV:1104.5499;%%

%\cite{Pugliese:2012ub}
\bibitem{Pugliese:2012ub}
  Pugliese D., Montani G. and Bernardini M. G.,
  %``On the Polish doughnut accretion disk via the effective potential approach,''
 Mon.\ Not.\ Roy.\ Astron.\ Soc.\ , {\bf 428},(2012) 2, 952-982
  %%CITATION = ARXIV:1206.4009;%%

\bibitem{Boy:1965:PCPS:}
R.~H. Boyer,
%\newblock {Rotating fluid masses in general relativity}.
\newblock { Proc. Cambridge Phil. Soc.}, 61:527, 1965.


\bibitem{Blaes1987}Blaes O.M.,  1987, MNRAS, 227, 975.


%\cite{Komissarov:2006nz}
\bibitem{Komissarov:2006nz}
  Komissarov S.~S.,
  %``Magnetized Tori around Kerr Black Holes: Analytic Solutions with a Toroidal Magnetic Field,''
  %Mon.\ Not.\ Roy.\ Astron.\ Soc.\
  MNRAS, {\bf 368} (2006) 993.
  %%CITATION = ASTRO-PH/0601678;%%

%\cite{Montero:2007tc}
\bibitem{Montero:2007tc}
  Montero P.~J., Zanotti  O., Font J.~A. and Rezzolla  L.,
  %``Dynamics of magnetized relativistic tori oscillating around black holes,''
 % Mon.\ Not.\ Roy.\ Astron.\ Soc.\
 MNRAS {\bf 378} (2007) 1101.
  %%CITATION = ASTRO-PH/0702485;%%


%\cite{Horak:2009iz}
\bibitem{Horak:2009iz}
  Horak J. and Bursa M.,
  ``Polarization from the oscillating magnetized accretion torus,'' in
   R. Bellazzini, E. Costa, G. Matt and G. Tagliaferri
 \emph{X-ray Polarimetry,
} Cambridge University Press
 2010, arXiv:0906.2420 .
  %%CITATION = ARXIV:0906.2420;%

 %%%%%%%%%%\alpha-\omega%%%%%%%%%%%%%%%%%%%%%%%%%%


 %\cite{Parker:1955zz}
\bibitem{Parker:1955zz}
  Parker E.~N.,
  %``Hydromagnetic Dynamo Models,''
  Astrophys.\ J.\  {\bf 122} (1955) 293.
  %%CITATION = ASJOA,122,293;%%

  %\cite{Parker:1970xv}
\bibitem{Parker:1970xv}
  Parker E.~N.,
  %``The Origin of Magnetic Fields,''
  Astrophys.\ J.\  {\bf 160} (1970) 383.
  %%CITATION = ASJOA,160,383;%%
  %
  \bibitem{Y.I.I2003}
    Yoshizawa A.,  Itoh S. I,  Itoh K.,
  \emph{Plasma and Fluid Turbulence: Theory and Modelling},
  CRC Press, 2003.





  \bibitem{R-ReS1999}
 Reyes-Ruiz  M., Stepinski T. F.,
 Astron.\ Astrophys.\  {\bf 342} (1999)  892–900.









\end{thebibliography}
\end{document}